\documentclass[useAMS,usenatbib]{mn2e}

\title[A tentative elementary model of quasars.]{ A tentative elementary model of quasars.}
\author[Jacques Moret-Bailly]{ Jacques Moret-Bailly$^{1}$\thanks{E-mail:
Jacques.Moret-Bailly@u-bourgogne.fr}\\
$^{1}$Laboratoire de Physique, Universit\' e de Bourgogne, 21000 Dijon France }
\begin{document}

\date{  }

\pagerange{\pageref{1}--\pageref{4}} \pubyear{2002}

\maketitle

\label{firstpage}

\begin{abstract}
A very simple model of quasar and Seyfert galaxy is obtained using the "Coherent Raman Effect on time-Incoherent 
Light" (CREIL) to explain a part of the observed redshifts. As its redshift is mostly provided by the CREIL, a QSO is 
not very far, its kernel may be a neutron star fed by the accretion of a disk, fed itself by the fall of many satellites. Disk 
and satellites are slowed by a relatively dense halo. The electric charges resulting of the friction of the disk produce 
flat lightnings, thus radio noise emitted mostly perpendicular to the disk, and X rays. The optical spectrum, including 
the shape of the spectral lines is explained without uncommon physics such as dark matter, by a splitting of the very 
wide lines resulting from simultaneous absorption and frequency shift, splitting due to a modulation of the redshift 
resulting from variations of a magnetic field.

\end{abstract}

\begin{keywords}
Quasars -- redshifts
\end{keywords}

\section{Introduction}

The standard explanation of the properties of the quasars and the Seyfert galaxies requires debatable hypothesis 
about the speed and the preservation of clouds; however, some observations, such as the radio-quietness of the BAL 
quasars remain unexplained.

Some astrophysicists looked for an alternative of Doppler effect to explain observed redshifts of light. But they failed 
because either they introduced strange, unknown matter, or they considered a Raman effect which blurs the images 
and the spectra.

The light scattering of a wide beam without frequency change may be coherent (in refraction and Bragg scattering), or 
incoherent (in Rayleigh scattering), but, with a frequency change by the scattering, only the incoherent scattering 
(regular Raman scattering) is usually studied. The remaining "white box" in the table of the light-matter interactions, is 
filled by the "Coherent Raman Effect on time-Incoherent Light" (CREIL) \citep{Mor98a,Mor98b,Mor01} whose 
principle is recalled in section 2. Rather than the history of CREIL, this section sets and answers the question: {\it 
how can a light-matter interaction imitate a Doppler effect?} The reader may skip this section.

Section 3 recalls the properties of the CREIL

The following section sets an elementary model which mixes the properties of the quasars and the Seyfert galaxies, 
introducing only simple concepts deduced from observation. It gives the origin of X rays, radio noise and their 
correlation with the BAL.

Section 5 explains the generation of the optical spectra.

\section{Finding "Coherent Raman Effect on time-Incoherent Light" (CREIL) as an alternative of Doppler effect.}

Suppose that a receiver observes a frequency lower than the frequency emitted by a coherent source, for instance 
that it observes three cycles while the source emits five. It remains two cycles more between the source and the 
receiver, so that the path is increased by two wavelengths: either the distance is increased, it is a Doppler (or 
expansion) effect, or the path is broken or curved, there is no more a good image. Thus, the incoherence of the light is 
necessary for an alternative of the Doppler effect which does not blur the images. \citet{Mar91} wrote that the 
incoherence is necessary, but he did not set that, consequently, {\it the incoherence must be a parameter of the 
searched effect,} so that the effect does not work with coherent light. The simplest introduction of incoherence, is 
setting that light is made of pulses whose length is of the order of five nanoseconds.

\medskip

Previous authors considered a scattering by a single molecule as in a Compton effect, while, in the optical domain, 
light interacts with sets of molecules, for instance in refraction and Bragg scattering.

Consider only wide beams, so that diffraction may be neglected. To deduce a new wave surface from a known one, 
Huyghens supposes that all points of the known wave surface are coherent sources of light which generate, after a 
short, given time, small wavelets whose envelope is the new wave surface. If molecules are placed on the initial wave 
surface, they scatter true wavelets, whose envelope gives a scattered wave surface nearly identical to a certain 
incident wave surface; as the number of molecules is finite, the generation of the envelope is not perfect, its flaws 
correspond to incoherently scattered light. If there is no frequency change, the interference of the scattered light 
which is late of $\pi/2$, decreases the phase of the incident beam, it is the refraction. The transfer of energy which 
produces a frequency change cancels the $\pi/2$ phase shift between the exciting and scattered beams {\it at the 
beginning of the excitation}; as this phase increases with the time, maintaining the time-coherence of the molecules to 
preserve the wave surfaces requires that the molecules do not restart by collisions: the scattered wave surface is 
identical to the incident one only if the length of the impulsions is shorter than the collisional time.

\medskip
Multiple Raman scatterings blur a spectrum, because each shifted frequency is partly scattered, so that the initial 
frequency remains beside frequencies resulting from many scatterings. Thus, avoiding a blur of the spectrum requires 
an interference of the scattered beam with the incident one into a single frequency beam.

This interference is geometrically possible because the incident and scattered wave surfaces are identical. A mixture 
of two frequencies cannot be observed by a spectrometer if the resulting amplitude is constant enough, that is if 
beats cannot be observed. At the start of a Raman scattering the exciting and scattered beams have the same phase 
so that the modulus of the elongations may be simply added provided that it does not appear a notable phase shift: 
the interference requires the equality of the initial phases and a time of observation, that is a length of the pulses, 
much shorter than the period of the beats. This explanation is weak, but it is verified by an elementary computation 
which shows also that the frequency shift of the incident light is proportional to the product of the relative Raman 
scattered amplitude by the Raman frequency and by the frequency of the incident light. Thus the relative frequency 
shift $\Delta\nu/\nu$ is almost independent on $\nu$.

\medskip
A spectral element in a beam, a mode, has a temperature given by Planck's law. Just as in refraction, the molecules do 
not perform fully transitions between the states connected by the Raman transition, they polarise. Acting as catalyst 
by their polarisations, the molecules transfer energy from a hot redshifted mode to a cold one which is blue-shifted; in 
fact at least all modes making the thermal radiation are blue-shifted, that is the thermal radiation is amplified.

\section{Properties of the CREIL}
The CREIL transfers by frequency shifts, energy from modes having the density of energy of a hot source to the 
thermal radiation, without a blur of the geometry of the beams. 

The relative frequency shift $\Delta\nu/\nu$ is nearly constant, it depends on the dispersion of the polarisability of 
the molecules which play the role of catalyst. Remark that supposing that the redshifts have a Doppler origin, some 
spectra of quasars appear distorted \citep{Webb}\footnote{Most references describe observations illustrating 
effects explained in the text.}. The dispersion of CREIL explains it without the hypothesis required by the standard 
theory: variation of the fine structure constant.

To be active in CREIL, a gas must have a collisional time longer than the length of the light pulses. The corresponding 
pressure depends on the gas, on the temperature, on a convenient definition of the collisions\dots Thus it is difficult 
to evaluate it; a rough order of magnitude is some Pascals. The gas must have low energy Raman transitions, in the 
radio-frequencies range. This is hyperfine transitions which may be between:

i) Genuine hyperfine levels if the molecule has nuclear non-zero spins. The H$_2^+$ molecule has probably 
convenient levels; as it self-destroys by collisions, its stability requires a very low pressure, thus working 
CREIL, so that its absorption lines are spread, mixed, invisible. An homogeneous density of undetectable 
H$_2^+$ may explain the Hubble redshift without expansion of the universe, and may be a source of 2.7K 
thermal radiation.

ii) Stark or Zeeman levels. Excited atomic hydrogen having a non-zero secondary $l$ quantum number works. The 
redshifting power of atomic hydrogen depends on the excitation of the atoms and on the fields. Magnetic fields are 
detected by Faraday rotations \citep{Welter}.

iii) Other levels in heavy mono- or poly-atomic molecules, uncommon in astrophysics.

\section{A model of quasars and Seyfert galaxies: non-spectroscopic properties}

In the standard theory, a quasar is surrounded by jets of mainly atomic hydrogen, and the line of sight goes through 
intergalactic clouds. Problems appear: conceiving the acceleration of the jets and the confinement of the clouds is 
difficult, requiring, for instance dark matter; the source of heating of the clouds is not clear; finally, the physico-
chemical parenthood, the similar density of jets and clouds having close redshifts is strange \citep{Tytler}.

A simultaneous absorption and redshift of a line produces a wide absorption line; a modulation of the redshift which 
may be done by a variation of a magnetic or electric field modulates the absorption, so that the wide lines appears 
split into pseudo-lines. Thus, suppose that the kernel is surrounded by a halo whose density and metallicity decrease 
with the distance to the kernel, and in which there are variable fields.

\medskip
The distance of the quasars is generally evaluated applying Hubble's law to the most redshifted lines; if it is applied 
to the less redshifted ones \citep{Petitjean,Shull}, the distance decreases much, so that the kernel of the quasar may 
be simply a neutron star.

It is often supposed that a quasar has an accretion disk. This disk is slowed by its friction against the halo, so that it 
falls, feeding the star with matter and thus with energy; as a disk is not homogenous, this feeding is irregular, so that 
the magnitude of the star changes slowly.

\medskip
Set $\theta \quad(0\leq\theta\leq\pi/2)$ the angle between the axis of the disk and the line of sight.

The friction of the disk electrifies it, as drops of rain in a cloud, but at a higher scale, with higher potentials and 
charges. The electric currents which appear as local lightnings are in the plane of the disk, thus flat, so that they 
radiate radiowaves as a flat homogenous antenna or a diffraction hole; thus the radio noise is maximum for $\theta=0$.

As the pressure is low, the disk works locally as a Crookes tube, X rays are emitted in the volume of the disk; for 
$\theta\approx\pi/2$, the X rays are partly absorbed while they propagate in the disk \citep{Brandt}: 
$\theta\approx\pi/2$ appears the condition for a BAL quasar.

We follow the analogy of quasars and Seyfert galaxies which is often outlined \citep{Weedman,Stein}; thus, it 
appears natural to suppose that the quasars have satellites as the kernels of the Seyfert galaxies, so that the main 
differences between these objects seem the number and the size of the satellites. The existence of a lot of satellites is 
backed by the observation of large decreases of magnitude during times of the order of the day: it may be 
occultations of the kernel by the satellites. The satellites may be planets rather than stars because in despite of their 
number they are not really visible, and the occultation of a neutron star does not require a big object. The satellites 
may be slowed by the halo, then broken into the disk, compensating the absorption of the halo by the kernel.

\section{Optical spectrum of quasars and Seyfert galaxies}
The halo may be divided into layers which give different types of spectra; starting from the kernel:

{\it Near the surface of the quasar.}

The pressure of the halo forbids a CREIL, but it is low enough to produces sharp emission then absorption lines, as 
an ordinary star.

\medskip
{\it The broad lines region.}

If a neutral atom of hydrogen has absorbed light at Lyman$_\alpha$ frequency, it returns to its ground state by a 
collision with an other atom or a free electron, heating the translational energy of the gas; the process is reversible, 
the hot gas (temperature larger than 10 000K) being able to emit a Lyman photon. These partly coherent interactions 
tend to reach an equilibrium between the translational energy of the gas and the mean temperature of the modes at the
resonance frequency.

The equilibrium is reached near the quasar for the strong lines, so that the intensity of light absorbed or amplified, at 
the strong resonance frequencies, corresponds to the temperature of the gas, not anymore to the temperature of the 
source. If the gas is slightly active for CREIL, the equilibrium is maintained during the frequency shift, so that a 
trough appears in the spectrum; small variations of the CREIL leave the light intensity constant in the trough, 
provided that the interaction is high enough to reach the equilibrium. The absorption does not depend on the column 
density and a lack of saturation does not need an explanation by a covering factor.

Leaving electric charges in the halo, the disk charges, so that its rotation induces a space-variable magnetic field close 
to it. This field may be increased by magnetic properties of the elements of the disk. The disk produces an electric 
field too. Thus, in some places close to the disk, hydrogen atoms pumped by Lyman transitions may be active for 
CREIL; where the field is high, a spectral element of the light resonates only along short path, so that it is not much 
absorbed; elsewhere the resonating spectral elements get the temperature of the gas.
Thus, if $\theta\approx\pi/2$, BEL and BAL appear and their correlation with the radio-quietness is clear.

Observing the quasar in radio-loud conditions ($\theta\ll\pi/2$), there is no magnetic field, no CREIL, the absorption 
which corresponds to the BAL is confused as an excess of absorption near the emission redshift 
\citep{Briggs,Anderson}. The propagation does not introduces a redshift, while it does in the other case; 
consequently the thermal radiation issued from the kernel is less amplified, the "dust emission" is lower than for BAL 
quasars \citep{Omont}.

\medskip
{\it The damped Lyman lines forest}

At a larger distance, the space-variable magnetic field necessary to split a wide Lyman line into pseudo-lines making 
the forest may be produced by the satellites.

The density of radiative energy is lower than close to the disk, that is the number of exciting modes coming from the 
kernel becomes low; the mean temperature of the exciting modes is low, so that the population in the excited states is 
almost negligible and the absorption may saturate.
If a magnetic field shifts the lines by CREIL the simultaneous absorption is low. The computation 
\citep{Mor01,Mor02} shows that the pseudo-lines may be sharp, but with strong feet, so that, by an increase of the 
absorption, they are damped more than a line having a collisional or Doppler profile.

\medskip
{\it The Lyman forest}

In the outer part of the halo, the lines of the forest may be sharp in despite of slow variations of the field because the 
absorption/redshift process produces a bistability: Suppose that a high field produces in a certain region a high 
CREIL; the spectral element whose frequency equals the Lyman frequency is permanently renewed, so that the atoms 
of hydrogen are strongly pumped and able to produce CREIL. Thus CREIL remains large until the field becomes very 
low, then, there is a brutal disappearance of CREIL. On the contrary, if CREIL is low, energy is not renewed at Lyman 
frequency, there are few excited atoms, a large magnetic field is needed to start a visible CREIL.

\medskip
A large fraction of the energy of the light is absorbed by the redshift and transferred to the thermal spectrum; it just 
happens that the most redshifted objects radiate a relatively hot thermal spectrum; is it produced by dust or by 
CREIL?

\section{Conclusion}
CREIL was found from a dependable theory which explains refraction and Bragg scattering; during a long time only its 
nonlinear version the "Impulsive Stimulated Raman Scattering" (ISRS) was commonly observed and used in the labs, 
so that it could appear careful to ignore CREIL. Now CREIL itself is observed \citep{Kandpal}, stronger than expected 
from a too cautious application of the theory, so that it must be taken into account.

The present simple model of quasars explains qualitatively more usual observations than the standard model, for 
instance the radio-quietness and the X absorption of the BAL quasars, and, however, requires only elementary 
physics. 

The model must be improved to become quantitative and adapted to all elements of the large family of quasars and 
Seyfert galaxies; it remains much work!

\end{document}